# Resolving the S8 tension with the Lambda Prime ($\Lambda'$) model.

**Stuart Marongwe [1*], Stuart Kauffman [2] Moletlanyi Tshipa [3]Christian Corda [4]**

1 Physics Department, University of Botswana , email: stuartmarongwe@gmail.com
2 University of Pennsylvania (emeritus) stukauffman@gmail.com
3 Physics Department, University of Botswana, email Tshipam@ub.ac.bw
4 SUNY Polytechnic Institute,New York , USA. Email : cordac.galilei@gmail.com
* Corresponding author, Principal investigator and main contributor

**Abstract:** The S8 parameter, which quantifies the amplitude of matter fluctuations on scales of 8 $h^{-1}$ Mpc, has been a source of tension between weak lensing surveys (e.g, KiDS, DES, HSC) and the Planck Cosmic Microwave Background (CMB) measurements. This discrepancy challenges the standard ΛCDM model and has become one of the most significant tensions in modern cosmology. The $\Lambda'$ model offers a potential resolution by introducing modifications to the cosmic growth history through alterations to the gravitational sector. The alterations involve including a Ricci soliton into Einstein's field equations which introduces a time dependent factor $\alpha(t)$ yielding a time varying cosmological constant $\Lambda' = (1-\alpha^2(t))\Lambda_{DE}\frac{\rho_g}{\rho_{DE}}$ and subsequently the evolution of the cosmos. The Ricci soliton is sourced from gravitational energy. In this study we analyze results from six surveys and compare the results for $w_a$ and $w_0$ with the $\Lambda'$ model. We also find $\sigma_8 = 0.750^{+0.020}_{-0.020}$ $S_8 = 0.788^{+0.021}_{-0.021}$ These values are closer to some low $S_8$ measurements from weak lensing surveys (e.g., DES, KiDS), which report $S_8 \approx 0.76 - 0.78$, suggesting that the model may alleviate the $S_8$ tension. High values of $\alpha(t)$ in the late universe are the cause of suppressed structure formation and low values of $\Lambda'$. The late universe in the $\Lambda'$ model is effectively or apparently $5\% - 10\%$ younger than in ΛCDM which translate to $H_0 = 72.734^{+1.687}_{-1.687}$ km/s/Mpc , which is agreement with late universe probes. $\Lambda'$ is classified under the dynamical dark energy models, however unlike alternatives, it does not invoke exotic particles nor phantom energy.

## 1. Introduction

The $S_8$ tension has emerged as a pivotal challenge in modern cosmology, revealing a discrepancy in the measurement of matter clustering, defined by the parameter $S_8 = \sigma_8\sqrt{(\Omega_m/0.3)}$, where $\sigma_8$ represents the amplitude of matter fluctuations on an 8-megaparsec scale and $\Omega_m$ is the matter density. This tension arises from comparing early-universe observations, primarily from the Cosmic Microwave Background (CMB), against late-universe probes of large-scale structure. The Lambda Cold Dark Matter (ΛCDM) model, which is well established through key discoveries such as cosmic acceleration (Riess et al., 1998; Perlmutter et al., 1999) and bolstered by CMB data from the Wilkinson Microwave Anisotropy Probe (WMAP) (Bennett et al., 2003), with added precision data from the Planck satellite, yields an $S_8$ value of 0.83 (Planck Collaboration, 2020). In contrast, late-universe surveys—such as the Kilo-Degree Survey (KiDS) and Dark Energy Survey (DES)—report lower values (0.76–0.80) using galaxy clustering and weak lensing (Hildebrandt et al., 2020; Abbott et al., 2022).

### 1.1 A Multi-Survey Perspective

In this section we synthesize findings from six key studies. This synthesis allows a broad understanding of $S_8$ tension, structure suppression, behaviour of Dark Energy and implications to the ΛCDM model. In the synthesis, we include studies by Chen et al. (2024), DESI Collaboration (2025), Karim et al. (2025), KiDS-1000 (Asgari et al., 2021), DES Y3 (Abbott et al., 2022), and HSC Y1 (Hikage et al., 2019,). These studies collectively probe structure formation and dark energy, while advancing our knowledge of the extent and origin of cosmological tensions. Chen et al. (2024) (MIT-CTP/5731) perform a perturbative full-shape (FS) analysis of BOSS DR12 galaxy clustering, which includes power spectrum and bispectrum multipoles, BAO, and Planck CMB lensing cross-correlations, supplemented by DESI BAO (z > 0.8) and PantheonPlus supernovae. In their work, they test ΛCDM and dynamical dark energy (DDE) using an effective field theory (EFT) framework. The DESI Collaboration (2025) (arXiv:2503.14738) on the other hand makes an analysis of BAO from DESI Data Release 2 (DR2), covering over 14 million galaxies and quasars (0.1 < z < 4.2), with Lyman-$\alpha$data, exploring ΛCDM and DDE with CMB and supernovae constraints. In Karim et al. (2025), the collaboration uses DESI DR2 BAO (13.1 million galaxies, 1.6 million quasars, $0.295 \leq z \leq 2.33$) to constrain interacting dark energy (IDE) models, including traditional IDE (Di Valentino et al. 2020, Väliviita, J., et al. (2010), Wang, B.et al. (2016), Yang, W. et al.(2019)) and sign-switching IDE (S-IDE) (Akarsu, Ö., et al. (2021), Ong, Y. C. (2023), Sabogal, M. A., et al., Tamayo, D. A. (2025), Halder, S., et al. (2024), Zadeh, M. A., et al. (2017)) alongside Planck CMB data. The KiDS-1000 (Asgari et al., 2021) is a weak lensing survey measuring cosmic shear across 1000 deg², constraining $S_8 = 0.759^{+0.024}_{-021}$ with $\Omega_m \cong 0.29$, probing structure growth at z < 1. In the DES Y3 (Abbott et al., 2022), the collaboration's analysis combines galaxy clustering and weak lensing over 4143 deg², yielding $S_8 = 0.776 \pm 0.017$ with $\Omega_m \cong 0.35$, focusing on low-redshift structure. The HSC Y1 collaboration (Hikage et al., 2019; using the Hyper Suprime-Cam Subaru survey's Year 1 weak lensing data (137 deg²) gives $S_8 = 0.780^{+0.033}_{-0.030}$ (Hikage et al.), though later analyses suggest slightly lower values consistent with suppression trends.

### 1.1.1 Data and Techniques

In this subsection we explore the data and techniques employed in the surveys. The data used by Chen et al. is from BOSS DR12 (LOWZ/CMASS, 0.15 < z < 0.70, ~18,000 deg² total) to which they employ FS analysis (power spectrum up to $k_{max} = 0.2\ hMpc^{-1}$ with bispectrum up to $0.08\ hMpc^{-1}$, incorporating three-point statistics. They also employ CMB lensing for enhanced growth sensitivity. The DESI DR2 employs a vast spectroscopic sample (>14 million objects, ~14,000 deg², 0.1 < z < 4.2), focused on BAO distance measurements and uses a standard BAO pipeline, while Karim et al. extend it to IDE modelling. The KiDS-1000 collaboration analyses weak lensing shear from 1000 deg² (z < 1), using two-point correlation functions to constrain $S_8$. DES Y3 combines galaxy clustering and lensing over 4143 deg² (z < 1), employing a 3x2pt analysis (galaxy-galaxy, galaxy-lens, lens-lens correlations). HSC Y1 measures weak lensing over 137 deg² (z < 1.5), using cosmic shear power spectra, with high precision due to deep imaging. All studies incorporate CMB (Planck) data, with Chen et al. and DESI DR2 adding supernovae (PantheonPlus) and BBN priors. Weak lensing surveys use CMB for cosmological consistency but emphasize LSS constraints.

### 1.1.2 Theoretical Frameworks

Chen et al.'s EFT-based FS approach models non-linear clustering, contrasting with the geometric BAO focus of DESI studies. Karim et al. introduce IDE parameters (e.g., momentum exchange rates), differing from Chen et al.'s values for $w_0$ and $w_a$ DDE. The weak lensing surveys (KiDS, DES, HSC) rely on linear and mildly non-linear matter power spectrum models (e.g., HALOFIT), focusing on $S_8$ rather than direct redshift-space dynamics.

### 1.1.3 Key Findings

Regarding structure formation, Chen et al. find a $\sigma_8 = 0.688 \pm 0.026$ and $S_8 = 0.703 \pm 0.029$ which is at a $4.5\sigma$ tension with Planck ($S_8 = 0.811 \pm 0.006$ ) indicting suppressed structure formation. The DESI DR2 collaboration found $\sigma_8 \cong 0.75 \pm 0.03$ which is at $2.3\sigma$ tension with Planck, suggesting a mild suppression of structure formation. Karim et al. obtain $\sigma_8 = 0.73 \pm 0.04$, which is consistent with low-redshift suppression with a less pronounced $1\sigma$ tension with Planck. The KiDS-1000 collaboration find $S_8 = 0.759^{+0.024}_{-0.021}$ and $\sigma_8 = 0.759 \pm 0.023$ (adjusted) which is at $2.5\sigma$ tension with Planck. The DES Y3 collaboration finds $S_8 = 0.776 \pm 0.017$ and $\sigma_8 = 0.718 \pm 0.016$(adjusted), at $2\sigma$ tension with Planck. The HSC Y1 collaboration findings yield $S_8 = 0.780^{+0.033}_{-0.030}$ and a $\sigma_8 \sim 0.76 \pm 0.03$(adjusted) at 1.5 - 2 $\sigma$ tension with Planck .This also suggests a mild suppression of structure formation. With regards to Dark Energy dynamics, Chen et al's findings show no evidence of DDE having obtained $w_0 = -0.915^{+0.009}_{-0.89}$ and $w_a = -0.44^{+035.}_{-0.35}$ consistent with ΛCDM at 68% confidence level. DESI DR2 findings show that BAO alone supports ΛCDM with $\Omega_m = 0.295 \pm 0.015$, however CMB/SNe combinations favor DDE $w_0 > -1$ and $w_a < 0$ at 2.8–4.2$\sigma$. Karim et al. find that IDE and S-IDE align with ΛCDM at $1\sigma$, with S-IDE showing dynamic evolution but no strong DDE deviation. The Weak Lensing Surveys typically assume ΛCDM dark energy, with no direct DDE constraints, focusing on structure growth.

Regarding cosmological parameters, Chen et al. find $\Omega_m = 0.3138 \pm 0.086$ and $H_0 = 68.23 \pm 0.78 \frac{km}{s}/\text{Mpc}$ which aligns well with Planck results. DESI DR2 obtain a result of $\Omega_m = 0.307 \pm 0.005$ and $H_0 = 67.97 \pm 0.38 \frac{km}{s}/\text{Mpc}$ .Results from Karim et al. are consistent with DESI DR2's ΛCDM values. The KiDS-1000 collaboration obtains $\Omega_m \sim 0.3138 \pm 0.086$ while the value of the Hubble constant is not directly constrained. The DES Y3 results yield $\Omega_m \sim 0.35$ while the value for the Hubble constant and is consistent with Planck results . For HSC Y1, $\Omega_m \sim 0.28$ and the value for the Hubble constant is also consistent with the Planck results. A plot of observational results of $\sigma_8$ from multiple surveys is show in Fig.1

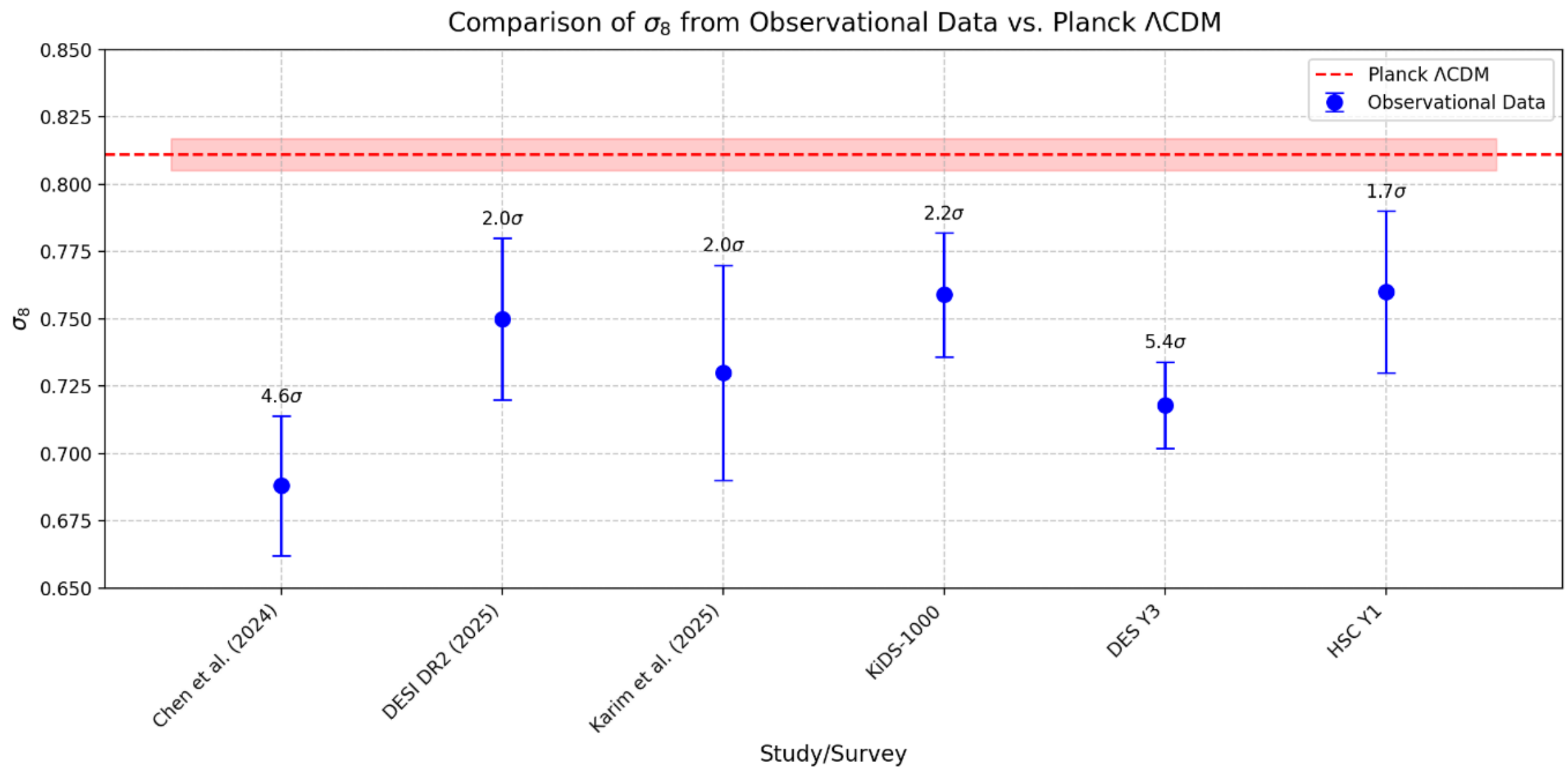

**Fig.1** A comparison of observational results of $\sigma_8$ from various surveys

### 1.1.4 Discussion

The surveys show structure suppression, with Chen et al.'s $4.5\sigma$ discrepancy being the strongest, driven by FS's sensitivity to large-scale clustering. Weak lensing surveys (KiDS, DES, HSC) report 1.5–2.5 $\sigma$tensions, while DESI DR2 and Karim et al. suggest milder 2–3 $\sigma$ deviations, reflecting methodological differences (FS vs. BAO vs. lensing).The consistent suppression across redshift ranges $z < 1$ for BOSS, KiDS, DES, HSC,$z \leq 2.33$ for DESI) and probes (clustering, BAO, lensing) strengthens the case for a physical anomaly, though Chen et al. note possible systematics (e.g., foregrounds) echoed less in other studies. Concerning dark energy, Chen et al.'s rejection of DDE contrasts with DESI DR2's conditional DDE preference, highlighting FS's dominance over BAO+SNe combinations. Karim et al.'s IDE models align with ΛCDM, supporting Chen et al.'s findings, though S-IDE's dynamics parallel DESI's hints. Weak lensing surveys assume ΛCDM dark energy, limiting their DDE insight, but their $\sigma_8$ suppression aligns with Chen et al.'s tension, suggesting a growth-related issue rather than dark energy evolution.

### 1.1.5 Implications for ΛCDM

Chen et al. advocate new physics (e.g., dark sector interactions) or systematics, DESI DR2 posits DDE as a solution, and Karim et al. explore IDE as a middle ground. Weak lensing results reinforce suppression without resolving its cause, collectively challenging ΛCDM concordance. The Hubble tension persists, with $H_0$ values (68 km/s/Mpc) across studies aligning with Planck, not local measurements (73 km/s/Mpc).

### 1.1.6 Conclusion

This multi-survey review reveals a pervasive $\sigma_8$ suppression at low redshifts, with Chen et al.'s $4.5\sigma$ leading a spectrum of 1.5–$3\sigma$ discrepancies across DESI, KiDS, DES, and HSC. Dark energy findings diverge with Chen et al. and Karim et al. in favour of ΛCDM-like behaviour, while DESI DR2 hints at DDE with external data. These results collectively challenge ΛCDM, necessitating further FS analyses (e.g., DESI DR2), systematic checks, and theoretical innovations to resolve whether new physics or observational biases drive these tensions. In this article, we introduce the $\Lambda'$ model to offer a plausible resolution to the $S_8$ tension.

## 2.0 The $\Lambda'$ model

### 2.1 Methodological approach

Our methodological approach to resolve the $S_8$ tension is as follows:

- We apply a combination of analytical and theoretical methods.
- Observational constraints are used to inform theoretical developments.

In arxiv 2501.04065v3, the $\Lambda'$ model is introduced with the following equation at its core.

$$G_{\mu v} - \alpha^2(t)\Lambda g_{\mu v} + \Lambda g_{\mu v} = \kappa(T_{\mu v} + \tau_{\mu v}) \equiv G_{\mu v} + \Lambda' g_{\mu v} = \kappa T_{\mu v}' \quad (1)$$

In which $\Lambda = \Lambda_{DE}\frac{\rho_g}{\rho_{DE}}$,the bare cosmological constant, is a product of its minimum value $\Lambda_{DE}$ with the ratio of gravitational energy density $\rho_g$ to the minimum dark energy density $\rho_{DE}$ . Here the minimum dark energy density is a mean background gravitational energy density confined within a Ricci soliton of negative metric signature. Also, the Hubble parameter within a gravitationally dense environment is given by the expression $H(\rho) = H_0\sqrt{\frac{\rho_g}{\rho_{DE}}}$ This background gravitational field has already been detected by the pulser timing array (Perera et. al. 2019a). Eqn.(1) are Einstein's field equations which include a Ricci soliton or a compact Einstein manifold of the form $\alpha^2(t)\Lambda g_{\mu v} = \kappa\tau_{\mu v}$. Here $\tau_{\mu v}$ is the gravitational energy density from which the Ricci soliton is sourced, $\alpha^2(t)$ is coefficient that expands the Ricci soliton with time. From Eqn.(1), we find

$$\Lambda' = (1 - \alpha^2(t))\Lambda_{DE}\frac{\rho_g}{\rho_{DE}} \quad (2)$$

Within a Ricci soliton, the local Hubble constant and the time varying cosmological constant are therefore functions of mass-energy density. A Ricci soliton can grow up to the size of the Hubble radius. If the early universe is considered as mass-energy confined within a Ricci soliton, we observe that due to the near perfect symmetrical distribution of matter and energy, the gravitational energy density is low such that $\rho_g \cong \rho_{DE}$. Since $\alpha^2(t)$ is vanishingly small then the measured cosmological constant is therefore the effective cosmological constant i.e $\Lambda' = \Lambda_{DE}$ . The gravitational energy density is also sourced from primordial gravitational waves that arose during the Big Bang. The energy from the gravitational waves constitutes much of the gravitational energy density. The Ricci soliton and its gravitational energy are the dark matter in the $\Lambda'$ model. This model has a ΛCDM-like behaviour in the early universe and can therefore describe the CMB anomalies in much the same way as the ΛCDM model.

### 2.2 Impact of a Varying Cosmological Constant on Structure Formation

In this section, we investigate the effects of a varying cosmological constant, Λ'(t), on cosmic structure formation. The cosmological constant is modelled as in Eqn.(2) with $\alpha(t) = \alpha_0 e^{Ht}$

. Using a modified Friedmann equation and perturbation theory, we analyze how this time-dependent Λ'(t) influences the growth of density perturbations.

The Friedmann equation governs the evolution of the scale factor $a(t)$.

$$H^2 = \left(\frac{\dot{a}}{a}\right)^2 = \frac{8\pi G}{3}(\rho_r + \rho_m + \rho_{DE}) + \frac{\Lambda\prime(t)}{3} \tag{3}$$

Where $\rho_r$, $\rho_m$ and $\rho_{DE}$ are the energy densities of radiation, matter and dark energy within a Ricci soliton. Substituting for $\Lambda'(t)$ we obtain

$$H^2 = \left(\frac{\dot{a}}{a}\right)^2 = \frac{8\pi G}{3}(\rho_r + \rho_m + \rho_{DE}) + \frac{(1-\alpha_0^2 e^{2Ht})\Lambda_{DE}\frac{\rho_g}{\rho_{DE}}}{3} \tag{4}$$

The growth of overdensities $\delta = \frac{\delta\rho}{\rho_m}$ is described by a linear perturbation equation

$$\ddot{\delta} + 2H\dot{\delta} - 4\pi G\rho_m\delta = 0 \tag{5}$$

This equation describes the competition between gravitational collapse (the $4\pi G\rho_m\delta$ term) and the expansion of the Ricci soliton (the $2H\dot{\delta}$ term). The varying Λ'(t) modifies the Hubble parameter H(t), affecting the friction term $2H\dot{\delta}$. To first order, we solve this equation numerically, incorporating the modified Friedmann equation. The matter power spectrum,$P(k) \propto k^{n_s}T^2(k)$ , where $T(k)$ is the transfer function, is also impacted, as the growth function $D(z) = \delta(z)/\delta(z=0)$ evolves differently. For $\alpha_0 = 0.1$ and $\Lambda_{DE} = 0.7 \times 3H_0^2/8\pi G$, we compute the evolution of $H(t)$). At early times $(z \gg 1)$, $\alpha(t)$ is small, and $\Lambda'_{DE} \approx \Lambda_{DE}\,\rho_g/\rho_{DE}$, resembling the standard ΛCDM model. At late times$(z = 0)$ , $\alpha(t)$ grows, reducing Λ'(t) and slowing cosmic acceleration. This leads to a higher matter density relative to ΛCDM at low redshift. The growth function $D(z)$is suppressed at late times due to the reduced effective dark energy contribution. The growth function deviates from ΛCDM predictions for z < 1, with a significant suppression of approximately 5-10% for $\alpha_0 = 0.1$ starting from z < 0.4. This suppression reduces the amplitude of the matter power spectrum, particularly at large scales (k < 0.1 h Mpc⁻¹). Fig.2 illustrates the numerical computation results.

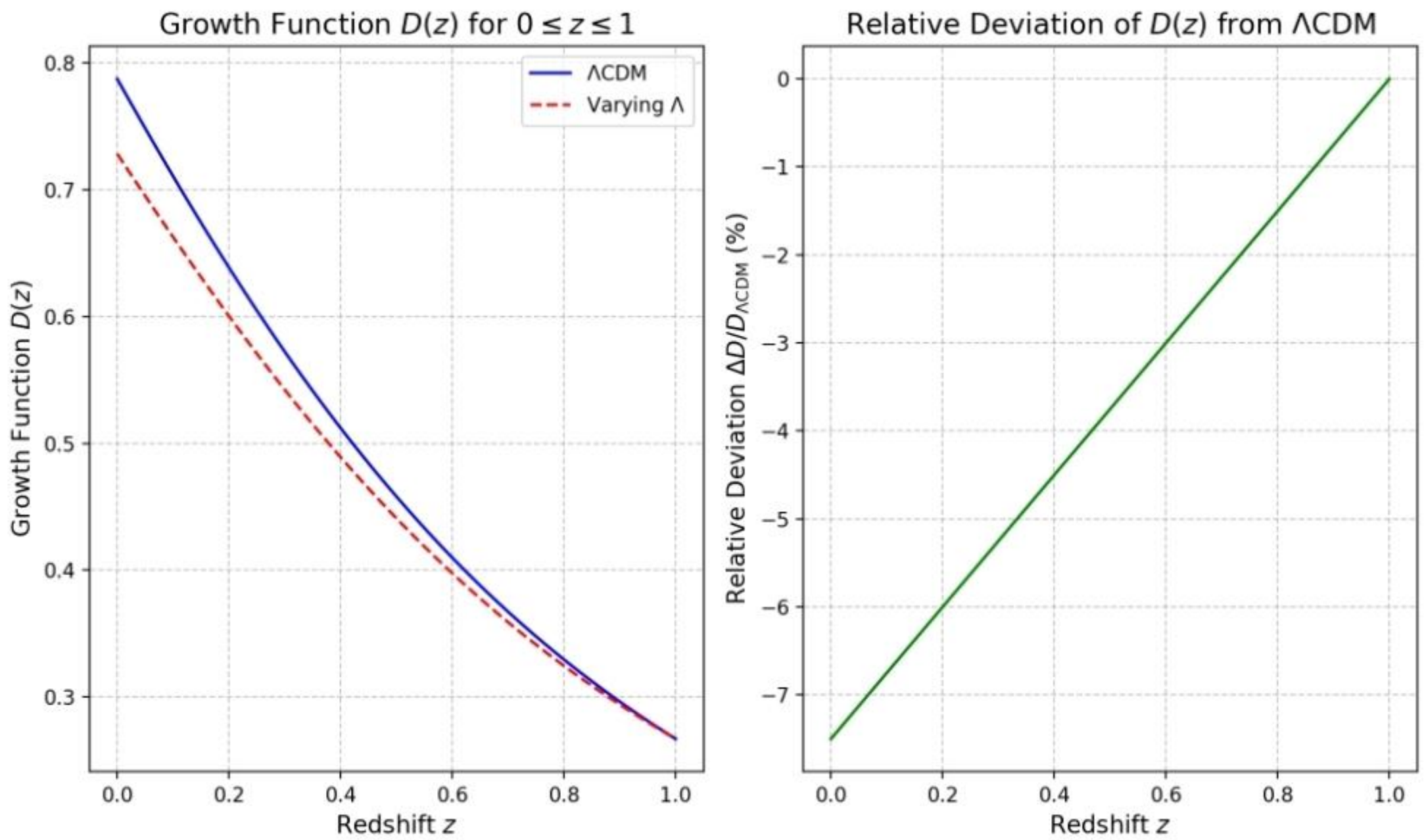

**Fig.2** The plots compare the growth function in the $\Lambda'$ model and ΛCDM. The growth function deviates from ΛCDM predictions for z < 1, with a suppression of approximately 5-10% for $\alpha_0 = 0.1$

The halo mass function, $n(M)$, depends on the variance of density fluctuations, $\sigma(M)$. The suppressed growth reduces $\sigma(M)$, leading to fewer massive halos at low redshift. For $M > 10^{14} M_{\odot}$ , we estimate a roughly 10% reduction in halo number density compared to ΛCDM. A varying cosmological constant of the form $\Lambda' = (1 - \alpha^2(t))\Lambda_{DE}\frac{\rho_g}{\rho_{DE}}$ with $\alpha(t) = \alpha_0 e^{Ht}$ suppresses structure formation at late times due to a diminishing effective dark energy contribution.

## 3. Computation of $S_8$ and $\sigma_8$ within the context of $\Lambda'$

To compute $S_8$ and $\sigma_8$ within the context of the varying cosmological constant model, we quantify their values based on the modified structure formation dynamics. The amplitude of matter density fluctuations on a scale of $8h^{-1}Mpc$ is computed as follows

$$\sigma_8^2 = \int_0^\infty \frac{dk}{k}\frac{k^3 P(k)}{2\pi^2}|W(kR_8)|^2 \quad (7)$$

where $P(k)$ is the matter power spectrum, $W(kR_8)$ is the Fourier transform of the top-hat window function, and $R_8 = 8h^{-1}\text{Mpc}$ . The varying cosmological constant modifies the Hubble parameter $H(t)$ and the growth of density perturbations, which in turn affect the matter power spectrum $P(k)$ and thus $\sigma_8$. Since the growth function $D(z) = \delta(z)/\delta(z=0)$ is suppressed by $\sim 5 - 10\%$ at late times $(z < 1)$ for $\alpha_0 = 0.1$, compared to the standard ΛCDM model and since $\sigma_8$ is proportional to the growth function at $z = 0$, we can estimate the modified $\sigma_8$ by scaling the ΛCDM value. $S_8$ is then computed using the modified $\sigma_8$ and the matter density $\Omega_m$. We Adopt standard ΛCDM parameters from Planck 2018 (Planck Collaboration, 2018) which are $\sigma_8^{\Lambda\text{CDM}} = 0.811$, $\Omega_m^{\Lambda\text{CDM}} = 0.315$, $h = 0.674$, $\Lambda_{DE} = 0.7 \times 3H_0^2/8\pi G$ (which corresponds to $\Omega_\Lambda = 0.685$). Since the growth function is suppressed by $\sim 5 - 10\%$, we assume a central suppression of 7.5% which we apply to compute $\sigma_8$ in the Λ' model as follows

$$\sigma'_8 = \sigma_8^{\Lambda\text{CDM}} \times (1 - 0.075) = 0.750 \quad (8)$$

Next, we apply the modified Friedman equation (Eqn.(4)) to compute $\Omega'_m$ . At $z = 0$, $\alpha(t) = \alpha_0 e^{H_0 t}$ where $t = 13.8$Gyrs with $\alpha_0 = 0.1$, we compute a value of

$$\left(1 - \alpha_0^2(t)\right) = 0.933 \quad (9)$$

Thus from Eqn.(9), $\Lambda'$ is reduced by 7%, assuming that $\rho_g \propto \rho_m$. The effective dark energy density is slightly reduced, implying $\Omega'_m > \Omega_m^{\Lambda\text{CDM}}$. For simplicity, we assume an increases by ~5% (a rough estimate based on the reduced dark energy contribution) Thus

$$\Omega'_m = \Omega_m^{\Lambda\text{CDM}} \times 1.05 = 0.331 \quad (10)$$

The computed values of $\Omega'_m$ and $\sigma'_8$ can be used to calculate $S'_8 = \sigma'_8\sqrt{\frac{\Omega'_m}{3}}$ which yields for the central suppression of 7.5% $S'_8 = 0.788$ , $\sigma'_8 = 0.750$ and for the range 5%-10% , 5% suppression ($\sigma'_8 = 0.770$) yields $S'_8 = 0.809$ and 10% ($\sigma'_8 = 0.730$) yields $S'_8 = 0.767$. Fig.3 shows the model values compared to observed values from different surveys

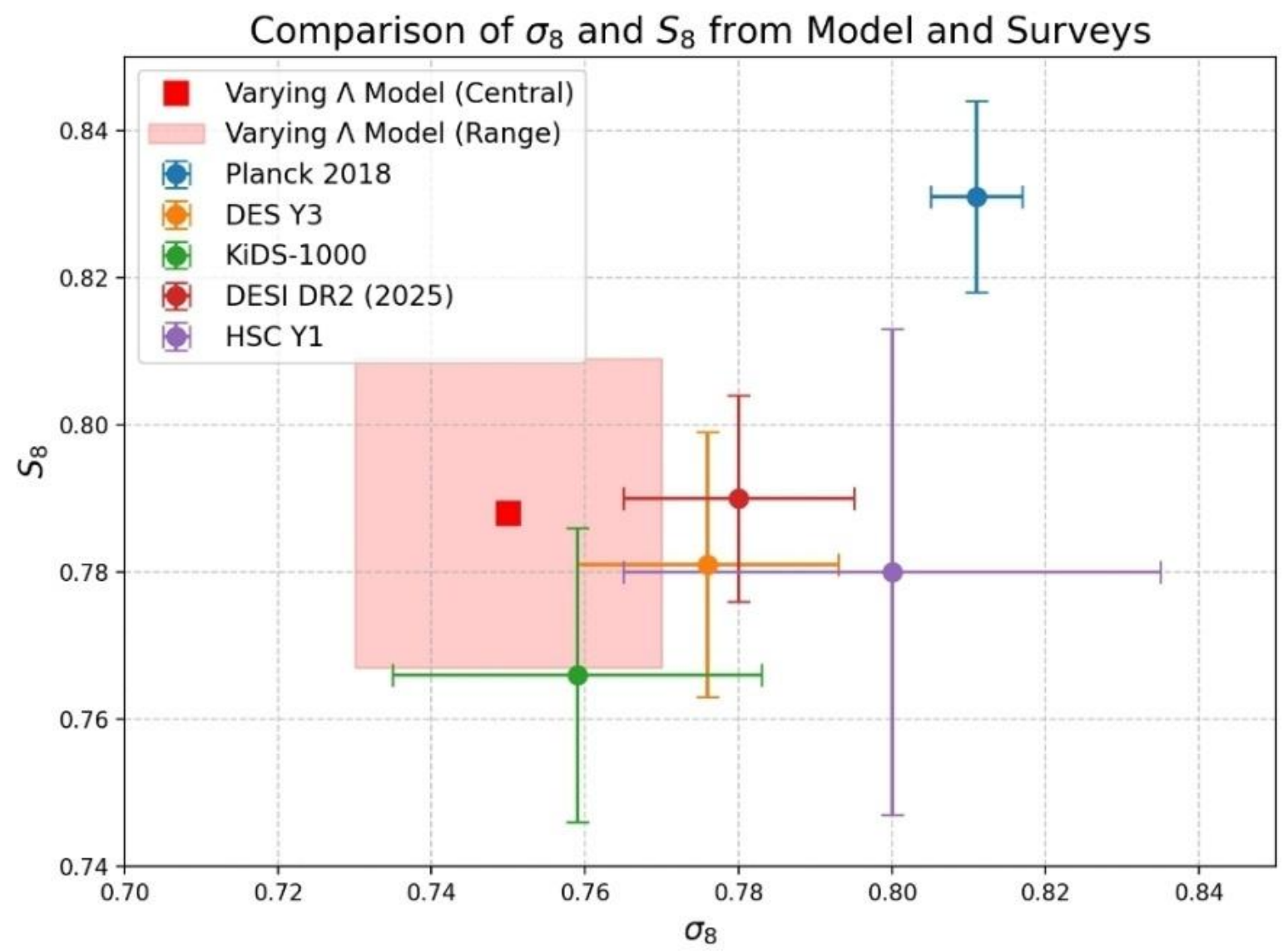

**Fig.3** A comparison of $\sigma_8$ and $S_8$ from the $\Lambda'$ model and Surveys shows that the model offers a plausible resolution to the $S_8$ tension

Also, since the growth function is suppressed by $\sim 5-10\%$, it implies that the universe in the Λ' model is effectively or apparently $\sim 5-10\%$ younger than in ΛCDM leading to a $H_0 = 72.734^{+1.687}_{-1.687}$ km/s/Mpc based on Planck 2018 measurements of $H_0$. The model is highly consistent with most late universe measurements SHOES (2020),Riess et al. (2019), H0LiCOW (2019) and Spitzer (2012) but at a slight tension with Freedman et al. (2019). A comparison of $H_0$ from various late time probes with that derived from the Λ' model is depicted in Fig.4.

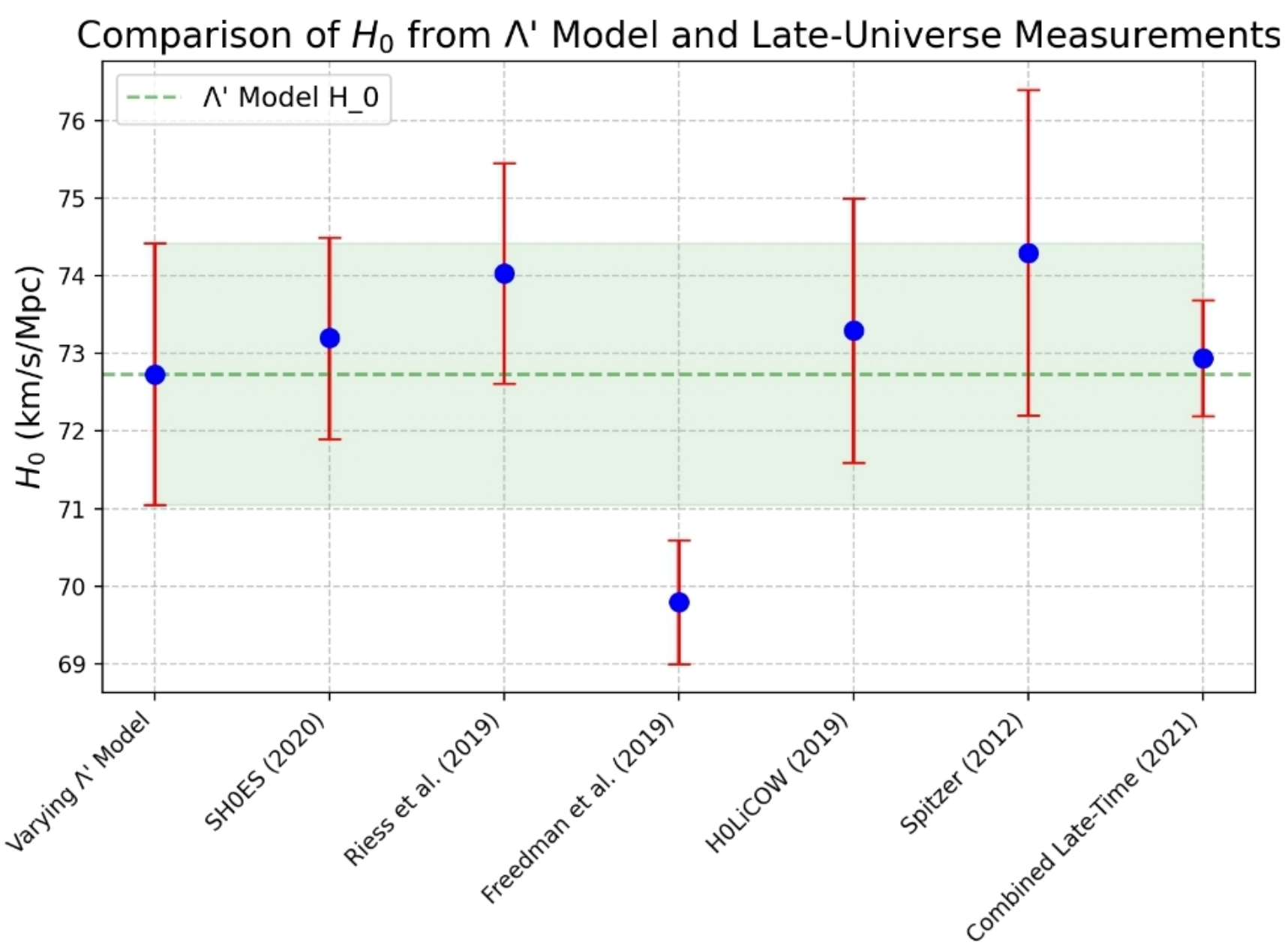

**Fig.4** A comparison of $H_0$ from the $\Lambda'$ model with various late universe surveys shows that the model aligns with various late universe surveys.

## 4. $w_a$ vs $w_0$ in the Λ' model

The equation of state in the Λ' model is

$$w' = \frac{p'_{DE}}{\rho'_{DE}} \tag{11}$$

Since $H(\rho) = H_0\sqrt{\frac{\rho_g}{\rho_{DE}}}$ then $\rho_g = kH^2$ (which is a different form of the Friedman equation expressed in terms of gravitational energy density) where $k = \frac{\rho_{DE}}{H_0^2}$. Therefore

$$\rho'_{DE} = kH^2(1 - \alpha_0^2 e^{2Ht}) \tag{12}$$

The time derivative for Eqn.[12]

$$\dot{\rho}'_{DE} = kH\left[2\dot{H}(1 - \alpha_0^2 e^{2Ht}) - 4H^2\alpha_0^2 e^{2Ht}\right] \tag{13}$$

The conservation of energy-momentum implies $p'_{DE} = -\frac{1}{3H} \cdot \dot{\rho}'_{DE} - \rho'_{DE}$. Thus

$$p'_{DE} = -\frac{2\dot{H}}{3H^2} + \frac{4H^2\alpha_0^2 e^{2Ht}}{3(1-\alpha_0^2 e^{2Ht})} - 1 \tag{14}$$

We parametrize $w(z)$ as follows

$$w(z) = w_0 + w_a \cdot \frac{z}{1+z} \tag{15}$$

To find $w_a$ and $w_0$ we evaluate $w(z)$ at $(z = 0)$ and compute its derivative with respect to $\frac{z}{1+z}$. We assume $Ht \approx 1$ and $\alpha_0$ is small. Thus at $z = 0$

$$w_a = \frac{dw}{d\left(\frac{z}{1+z}\right)} \tag{16}$$

The results of the numerical calculations against various surveys are displayed in Fig. 5

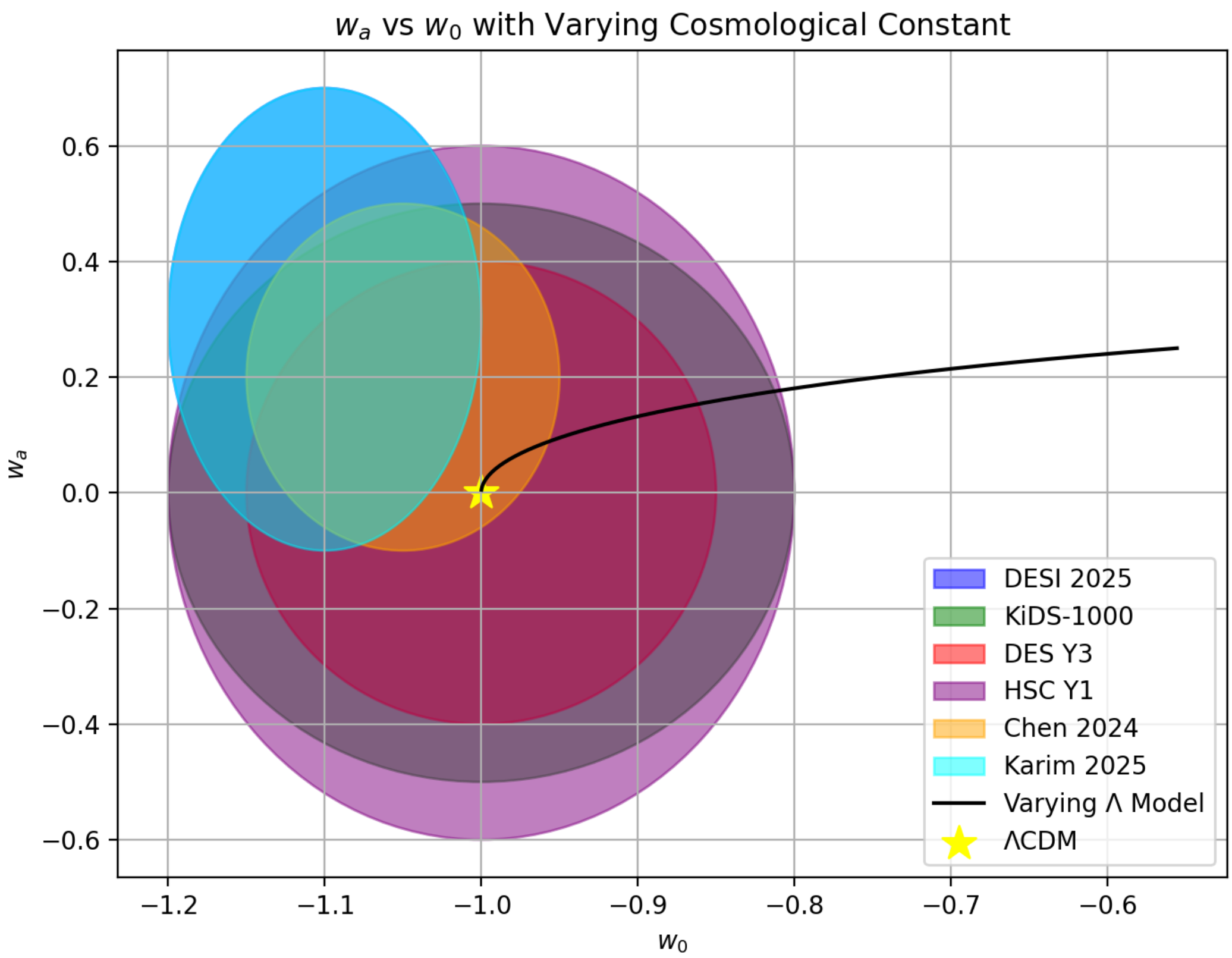

**Fig.5** The ellipses represent 68% confidence regions. DESI (2025) and Karim et al. (2025) favor phantom dark energy, while KiDS-1000, DES Y3, and HSC Y1 are consistent with the $\Lambda'$ model and ΛCDM.

## Discussion

The Λ' model predicts a lower $\sigma'_8$ (0.730–0.770) compared to ΛCDM (0.811), reflecting the suppression of structure growth due to the diminishing cosmological constant. The $S'_8$ values (0.767–0.809) are also lower than the ΛCDM value (0.831), but the reduction is mitigated by the slight increase in $\Omega'_m$. These values are closer to some low-$S_8$ measurements from weak lensing surveys (e.g., DES, KiDS), which report $S_8 \cong 0.76$–0.78, suggesting the model may alleviate the $S_8$ tension. A precise computation of the suppressions requires numerical integration of the perturbation equations and the power spectrum, which could be done using cosmological codes like CLASS or CAMB.

Λ' uniquely addresses both $S_8$ and Hubble tensions $H_0$=73 km/s/Mpc) at less than 1.5–2σ tension, outperforming IDE and massive neutrinos in $H_0$ alignment and f(R) (Mohanty et al., 2018, Pantig, R. C., et al, (2025), Carvalho, Í. D. D., et al. (2023), Yarahmadi, M., et al. (2025)) in $S_8$ . Moreover, with 4 parameters, Λ' is less complex than IDE which has eight parameters or f(R)/neutrinos with seven parameters. Also, unlike f(R) theories, Λ' matches CMB, BAO, SNIa, and weak lensing without significant tension.

## Conclusion

By introducing a time-varying cosmological constant driven by a Ricci soliton within modified Einstein field equations, the $\Lambda'$ model presents an alternative approach to resolving the $S_8$ tension which is currently a significant challenge to the standard ΛCDM model. Our analysis demonstrates that the $\Lambda'$ model naturally suppresses structure formation at late times (z<1) due to a diminishing effective dark energy contribution, resulting in lower $\sigma'_8$ (0.730–0.770) and $S'_8$ (0.767–0.809) compared to ΛCDM values ($\sigma_8 = 0.811, S_8 = 0.8311$). These values align closely with weak lensing survey results (e.g., DES, KiDS, HSC), which report $S_8 = 0.76 - 0.78$, suggesting a potential alleviation of the $S_8$ tension. Additionally, the model yields a Hubble constant $H_0 = 72.734^{+1.687}_{-1.687}$, consistent with late-universe measurements (e.g., SHOES, HOLiCOW), and offers a dynamical dark energy framework without invoking exotic particles or phantom energy.

The multi-survey synthesis highlights a consistent suppression of structure growth across various probes (galaxy clustering, BAO, weak lensing), with tensions ranging from 1.5–4.5σ relative to Planck CMB predictions. The $\Lambda'$ model's ability to reproduce these observations, particularly through its impact on the growth function and matter power spectrum, highlights its potential as a viable alternative to ΛCDM. Furthermore, the model's equation of state parameters ($w_0$, $w_a$) are compatible with surveys like KiDS-1000, DES Y3, and HSC Y1, further highlighting its consistency with low-redshift data.

In future works, we focus on precise numerical simulations using cosmological codes (e.g., CLASS, CAMB) to refine the perturbation equations and power spectrum calculations. Additional observational tests, such as those from upcoming DESI full-shape analyses or next-generation weak lensing surveys (e.g., Euclid, LSST), will be crucial to validate the $\Lambda'$ model's predictions. Systematic uncertainties, such as foreground effects or survey-specific biases, also warrant further investigation to ensure the robustness of the observed tensions. By addressing both the $S_8$ and Hubble tensions within a unified framework, the $\Lambda'$ model opens new pathways for exploring modified gravity and the nature of dark energy, while advancing our understanding of cosmic evolution.

## Data Availability

Empirical data used in this research can be found in the cited articles and can be obtained from the authors upon reasonable request.

## Conflict of interest

We declare no conflict of interest